\documentclass[aps,prd,twocolumn,superscriptaddress,amssymb,eqsecnum,showpacs,showkeyes,secnumarabic,graphics,floatfix,nofootinbib,tightenlines,longbibliography]{revtex4-1}

\usepackage{graphicx}
\usepackage{epsfig}
\usepackage{bm}
\usepackage{cancel}
\usepackage{dcolumn}
\usepackage{amsmath}
\usepackage{hhline}
\usepackage[utf8]{inputenc}
\usepackage[breaklinks=true,colorlinks=true,
linkcolor=blue,urlcolor=blue,citecolor=cyan,% PDF VIEW
bookmarks=true,bookmarksopenlevel=2]{hyperref}

\def \beq  {\begin{equation}}
\def \eeq  {\end{equation}}
\def \ber  {\begin{eqnarray}}
\def \eer  {\end{eqnarray}}

\begin{document}
\newcommand{\newc}{\newcommand}

\newc{\be}{\begin{equation}}
\newc{\ee}{\end{equation}}
\newc{\ba}{\begin{eqnarray}}
\newc{\ea}{\end{eqnarray}}
\newc{\bea}{\begin{eqnarray*}}
\newc{\eea}{\end{eqnarray*}}
\newc{\D}{\partial}
\newc{\ie}{{\it i.e.} }
\newc{\eg}{{\it e.g.} }
\newc{\etc}{{\it etc.} }
\newc{\etal}{{\it et al.}}
\newc{\lcdm}{$\Lambda$CDM}
\newcommand{\nn}{\nonumber}
\newc{\ra}{\Rightarrow}

\date{\today}
\title{Propagation of gravitational waves in an expanding background in the presence of a  point mass}

\author{I. Antoniou}\email{ianton@uoi.gr}
\affiliation{Department of Physics, University of Ioannina,
GR-45110, Ioannina, Greece}
\author{D. Papadopoulos}\email{papadop@astro.auth.gr}
\affiliation{Department of Physics, University of Thessaloniki,
Thessaloniki,54124, Greece}
% \affiliation{Department of Physics,
%University of Ioannina, GR-45110, Ioannina, Greece}
\author{L. Perivolaropoulos}\email{leandros@uoi.gr}
\affiliation{Department of Physics, University of Ioannina,
GR-45110, Ioannina, Greece}

\begin{abstract}
We solve the Laplace equation $\Box h_{ij}=0$ describing the propagation of gravitational waves in an expanding background
metric with a power law scale factor in the presence of a point mass in the weak field approximation (Newtonian McVittie background). We use boundary conditions at large distances from the mass corresponding to a standing spherical gravitational wave in an expanding background which is equivalent to a linear combination of an incoming and an outgoing propagating gravitational wave. We compare the solution with the corresponding
solution in the absence of the point mass and show that the point mass increases the amplitude of the wave and also decreases its
frequency (as observed by an observer at infinity) in accordance with gravitational time delay.

\end{abstract}
\pacs{04.30.Nk,04.30.-w}
\maketitle

\section{Introduction}
\label{sec:Introduction}

The theoretical prediction of gravitational waves (GWs) originates
in 1893 when Heaviside first discussed the possibility of their
existence. In 1916 Einstein predicted their existence
\cite{Einstein:1916cc,Einstein:1918btx} in the context of general
relativity. In the linearized weak-field approximation, he found
that his equations had transverse wave solutions travelling at the
speed of light \cite{Ellis:2016rrr} produced by the time
dependence of the mass quadrupole moment of the source
\cite{LeTiec:2016sgy}. Einstein realized that GW amplitudes would
be small and up until 1957, there was debate about the physical
relevance of their existence \cite{Saulson:2010zz}.\

Nevertheless, the discovery of the binary pulsar system PSR
B1913+16 by Hulse and Taylor \cite{Hulse:1974eb} and subsequent
observations of its energy loss by Taylor and Weisberg
\cite{Taylor:1982zz} demonstrated indirectly the existence of GWs.
This discovery, along with subsequent related analysis
\cite{Press:1972am}, led to the recognition that a possible direct
detection and analysis of GWs could reveal interesting properties of various relativistic systems and could also provide new tests of
general relativity, especially in
the strong-field regime.\\

Recently, Abbot et al
\cite{TheLIGOScientific:2016agk,Abbott:2016blz}, reported the
first direct detection of GWs emitted by
a binary black hole (BBH) system merging to form a single black hole (BH).
Their observation provides a direct window to the properties of
space time in the strong-field limit and is consistent with
predictions of general relativity for the nonlinear dynamics of
highly disturbed BHs. The announced beautiful discovery is the
result of great efforts for a century by several scientists
\cite{Abbott:2016blz} (and references therein). It is a great
investigation, because we have now one more window to the Universe
and one more confirmation of the theory of general relativity.
Such GW observations can be used to test the equivalence principle
\cite{Kahya:2016prx,Wu:2016igi,Liu:2016edq,Yunes:2016jcc}, test
the propagation of GWs
\cite{Yunes:2016jcc,Branchina:2016gad,Schreck:2016qiz,Arzano:2016twc,Bicudo:2016pps,Collett:2016dey,Blas:2016qmn,Garcia-Bellido:2016zmj},
test the validity of general relativity
\cite{Konoplya:2016pmh,Moffat:2016gkd,Vainio:2016qas}, constrain
early cosmological phase transitions
\cite{Dev:2016feu,Jaeckel:2016jlh}, probe the quantum structure of
black holes \cite{Giddings:2016tla}
or the connection between dark matter and primordial black holes \cite{Sasaki:2016jop,Bird:2016dcv,Clesse:2016vqa}. \\

The recent direct discovery of the GWs has been achieved by the
LIGO/Virgo collaboration associating the GW 150914 event
\cite{TheLIGOScientific:2016agk,Abbott:2016blz,Maselli:2016ekw}
to the coalescence of a BBH. This binary detection suggests that
BBH masses and merging rates may be higher than estimated
previously. The rates however, are in agreement with more recent
estimates, obtained with a population synthesis approach,
predicting the early formation of detectable BBH
\cite{Dominik:2012kk,Dominik:2014yma}. Thus, the
stochastic gravitational waves background (SGWB) produced by merging cosmological BBH sources
could be larger than previously assumed \cite{Maselli:2016ekw} (and references therein) and
may be detectable by advanced detectors
\cite{TheLIGOScientific:2016wyq}.\\

A
stochastic background of relic gravitational waves (RGWs) is predicted by inflationary models  \cite{Nakama:2015nea,Nakama:2016enz} and
has been well studied \cite{Starobinsky:1979ty,Abbott:1985cu}. The power spectrum of
relic gravitational wave background reflects the physical
conditions in the early Universe thus providing valuable
information for cosmology \cite{Dufaux:2007pt}. This spectrum
is determined by the early stage of inflation as well as by the expansion properties of the subsequent
epochs, including the current one. The calculation of the spectrum
\cite{Allen:1987bk,Henriques:2003ga} was initially performed for a
currently decelerating universe. However, it is now well known
that the universe expansion is currently accelerating
\cite{Riess:1998cb,Perlmutter:1998np} and since the evolution of
RGWs depends on the expanding background space time,  the spectrum of RGWs should be modified accordingly. This
modification was confirmed and studied in Refs
\cite{Zhang:2005nw,Izquierdo:2004gk} using the well-known
formulation of GWs in an expanding Universe
\cite{Grishchuk:1974ny} and an approximation of the scale
factor $a(\tau)$ in the context of a sequence of successive expansion epochs,
including the current stage of accelerating expansion. It was
found that the current accelerating expansion induces
modifications in both the shape and the amplitude of the RGW
spectrum.

Since existence of RGWs is a key prediction of the
inflationary models, their detection could provide
 evidence that inflation actually took place. Thus,
it is important to accurately calculate the expected detailed form
of the RGW spectrum. Calculations related to RGWs in an
accelerating Universe have been performed \cite{Zhao:2006eb} and a
numerical method has been developed to calculate the power
spectrum of the RGWs. Late evolution of RGWs in coupled dark
energy models has been examined extensively in Ref. \cite{Almazan:2013bea}.\\

Even though the effects of cosmological expansion on GWs have been
investigated mainly in the context of RGWs, these effects are
relevant in all cases when the source is located at cosmologically
large distances from the observer (redshift $z\geq 0.1$). The GW
150914 ($z=0.09$) event is in the limit of such distances and
therefore, the effects of cosmic expansion may be relevant.  Thus
a wide range of studies have investigated the effects of
cosmological expansion of GWs from a variety of viewpoints
including the effects of expansion on the GW group and phase
velocities  \cite{Balek:2007hh,Brillouin}, mathematical aspects
and exact solutions
\cite{Fabris:1998ak,Alekseev:1995uw,Tauber:1985yp,Waylen:1978dx,Tamayo:2015qla},
quantum and thermodynamic properties of GWs
\cite{Arzano:2016twc},\cite{IzquierdoSaez:2005ja}, general
properties \cite{Hartnett:1900zzc,Waylen:1978dx,Jackson:2008yv}
nonlinear effects \cite{Ikeda:2015hqa}, properties of the GW
energy momentum tensor \cite{Su:2012hf}, collision of GWs with
electromagnetic waves
\cite{Alekseev:2015wbn}, evolution of GWs in gravitational plasma \cite{Baptista:1982dc} etc .\\

Even though these studies have properly taken into account the
expansion of the background metric, they have not taken into
account the effects of the gravitational field of  mass
distributions on the evolution of the GWs. Such a gravitational
field combined with the expanding background may induce new
observable effects on the spectrum of propagating gravitational
waves affecting the amplitude and the frequency of such waves (due
to gravitational time delay) \cite{Zeldovich:1992if}.

Assuming spherical symmetry, the background metric around a point
mass embedded in an expanding Friedmann-Lemaitre-Robertson-Walker(FLRW) cosmological background
is well approximated by the McVittie \cite{McVittie:1933zz,9780511535185} spacetime. Such a metric is
further simplified in the Newtonian limit and has been used as the background metric for the investigation of bound system geodesics
in phantom and quintessence cosmologies \cite{Nesseris:2004uj,Antoniou:2016obw,Faraoni:2007es,Nolan:2014maa}.
This metric can also be used as a background for the propagation of GWs in order to investigate the influence of a mass
distribution of a GW propagating in an expanding cosmological background.

In the present analysis we address the following question: 'What
are the weak field effects of a point mass on a multipole spherical wave
component of a GW evolving in an expanding background in the
vicinity of the mass?'. In particular we numerically solve the
dynamical equation for the evolution of GWs in the background of
the Newtonian McVittie metric and identify the effects induced by
the point mass on the amplitude and frequency of the evolving
GW as a function of the parameters determining the mass for fixed
background expansion rate. As a test of our analysis, in the zero
mass limit, our numerical solution reduces to the well known
analytic solution of a GW evolving in an expanding background.

The structure of this paper is the following: In the next section
we review the wave equation and the behavior of the GW in a
homogeneous-isotropic expanding background. We also derive the
gravitational wave equation in a background metric corresponding
to the Newtonian limit of the McVittie metric. In section $III$ we
solve numerically the wave equation in the Newtonian McVittie
background and identify the new features induced in the GW by the
presence of the point mass. Finally in section $IV$ we conclude,
summarize and discuss possible extensions on this analysis.

\section{Gravitational Waves in Expanding Universe in the Presence of Point Mass}
We first briefly review the propagation evolution of a plane GW in the
$\hat{z}$ direction (the direction of the wavevector $\vec{k}$)
with tensor perturbations in the $x-y$ plane. The perturbations to
the metric are described by two functions, $h_+$ and $h_\times$,
assumed small. We use the FRW metric in cartesian coordinates with
the components $g_{00}=1$, zero space-time components $g_{0i}=0$,
 and set $c=1$. The spatial part of the
metric is of the form: \be g_{ij}= -a^2(t)\begin{pmatrix}
  1+h_+ & h_\times & 0 \\
  h_\times & 1-h_+ & 0 \\
  0 & 0 & 1\\
\end{pmatrix}
\label{metric1}\ee

The perturbation tensor $H_{ij}$ is symmetric, divergenceless,
traceless and has the form:  \be H_{ij}= \begin{pmatrix}
  h_+ & h_\times & 0 \\
  h_\times & -h_+ & 0 \\
  0 & 0 & 1\\
\end{pmatrix}
\label{tensor}\ee From the Einstein equations for tensor
perturbations, it is easy to derive a set of equations governing
the evolution of the tensor variables $h_+$ and $h_\times$. We
write the FRW metric in cartesian coordinates and in conformal time $\tau$  (defined by $dt=a d\tau$), in the form \be
ds^2=a^2(\tau)[d\tau^2-(\delta_{ij}+H_{ij})dx^{i}dx^{j}]
\label{metrconf}\ee The dynamical equation determining the
evolution of the GWs is of the form: \be
 \Box H_{ij}= \partial_{\mu}(\sqrt{-g}\partial^{\mu}H_{ij}(\vec{r},\tau)) =0 \label{waveeq}\ee
Since all components of the tensor perturbations evolve in accordance with the same wave equation (\ref{waveeq}) we may set $H_k\equiv H_{ij}$.
Without loss of generality we assume propagation in the $z$
direction and thus we use the ansatz:
 \be  H_{k}(\tau,z)=h_{k}(\tau) e^{\pm ikz} \label{waveans}\ee  Using eq. (\ref{waveans}) in (\ref{waveeq}) we find the dynamical equation for the evolution of gravitational waves in conformal time in an FRW background as
%~\cite{Schluessel:2014nwa}
\be h''_{k}+2\frac{a'}{a}h'_{k}+k^2h_{k}=0 \label{waveeq1}\ee
where the prime $'$ denotes the derivative with respect to
conformal time. Notice that  all of the perturbation tensor
components obey the same equation. We introduce a rescaling of
conformal time as $\bar{\tau}=k\tau$ and thus it becomes clear
that \be h_k(\tau)=h(k\tau) \label{hresc} \ee The rescaling
expressed by eq. (\ref{hresc}) can only be made in conformal time
provided that the scale factor is a power law $a(\tau)\sim
\tau^\alpha$. In the radiation dominated epoch we have $\alpha=1$
and during the matter dominated era $\alpha=2$. The wave solution
(\ref{waveans}) can be written in spherical coordinates as: \be
H_k(\tau,\rho,\theta)=h(k\tau)e^{\pm ik\rho\cos\theta}
\label{hsphercoord}\ee The spectrum of the GWs may be obtained as
~\cite{Zhang:2006mja}: \be P(k,\tau)=\frac{4\; l_{Pl}}{\sqrt{\pi}}k
\mid h(k\tau)\mid \label{powersp}\ee
where $l_{Pl}\sim \sqrt{G}$ is the Planck length.
The plane wave of equation (\ref{hsphercoord}) can be expanded in
spherical waves as: \be e^{i k \rho\cos{\theta}}=\sum_{l=0}^{\infty}i^l(2l+1)j_{l}(k\rho)P_l(\cos{\theta})\label{planeexp}\ee where $j_l(x)$ are the
spherical Bessel functions and $P_l(\cos{\theta})$
are Legendre's polynomials. Thus, the partial spherical GW is (at
order $l$) \be H_{ij}(\tau,\rho,\theta) \sim
h(k\tau)j_{l}(k\rho) P_l(\cos{\theta})\label{planeordl}\ee

After rescaling, the dynamical equation (\ref{waveeq1}) is written
as: \be
 h''(k\tau)+2\frac{a'}{a}h'(k\tau)+h(k\tau)=0 \label{rescwaveeq}\ee
where the prime $'$ now denotes differentiation with respect to
the rescaled conformal time $k\tau$.

Assuming a power law for the background scale factor as
$a(\tau)\sim \tau^\alpha$, the solution of the wave equation
(\ref{rescwaveeq}) can be written in terms of incoming and outgoing waves (Hankel functions) as \be
h(k\tau)=\frac{1}{a(\tau)}(\tilde{A}_{k}\sqrt{k\tau}H^{(1)}_{\alpha-1/2}(k\tau)+\tilde{B}_{k}\sqrt{k\tau}H^{(2)}_{\alpha-1/2}(k\tau))\label{wavesol1}\ee
where $H^{(1)}$, $H^{(2)}$ are the Hankel functions and
$\tilde{A}_{k}$, $\tilde{B}_{k}$ are arbitrary constants which may
depend on $k$ and are determined by the initial conditions. This
solution may also be written in terms of standing waves as \be
h(k\tau)=(k\tau)^{\frac{1}{2}-\alpha}(A_{k}J_{\alpha-1/2}(k\tau)+B_{k}Y_{\alpha-1/2}(k\tau))\label{wavesol2}\ee
where $J$,$Y$ are the Bessel functions of first and second kind
respectively and $A_k=\tilde{A}_k k^\alpha$, $B_k=\tilde{B}_k
k^\alpha$. Thus, for a power law scale factor the spherical GW is
\begin{eqnarray}
&&H_k(\tau,\rho,\theta)=(k\tau)^{\frac{1}{2}-\alpha}(A_{k}J_{\alpha-1/2}(k\tau)+B_{k}Y_{\alpha-1/2}(k\tau))\nonumber\\&&j_{l}(k\rho)P_{l}(cos\theta)\label{wave}\end{eqnarray}

As a warm up exercise before the introduction of a point mass in
the metric, we now rederive the solution (\ref{wavesol2}),
(\ref{wave}) starting from the FRW metric in spherical coordinates
\begin{equation}\label{frwsphercoords}
ds^2=a(\tau)^2\{d\tau^2-[d\rho^2+\rho^2(d\theta^2+\sin^2{\theta}d\phi^2)]\}
\end{equation}

Because of the azimuthal symmetry, the solution doesn't depend on
the variable $\phi$ we will seek solutions of the Eq.
(\ref{waveeq}) of the form
\begin{equation}\label{waveansspher} H_k(\tau,\rho,\theta,\phi)=
f(\tau)R(\rho)P_{l}(cos\theta) \end{equation} From the Eqs.
(\ref{waveeq}),(\ref{waveansspher}), after separation of variables
we find
\begin{equation}\label{rdiffeqsep}
\frac{1}{R}[\frac{d^2 R}{d \rho^2}+\frac{2}{\rho}\frac{d R}{d
\rho}-\frac{l(l+1)}{\rho^2}]=-k^2
\end{equation}
and
\begin{equation}\label{fdiffeqsep}
\frac{1}{f}[\frac{d^2 f}{d \tau^2}+\frac{2a'}{a}\frac{d f}{d
\tau}]=-k^2
\end{equation}
where $k^2$ is arbitrary constant. As expected eq.
(\ref{fdiffeqsep}) is identical to eq. (\ref{waveeq1}) while eq.
({\ref{rdiffeqsep}) is the spherical Bessel equation  with
acceptable solution
\begin{equation}\label{reqjl}
R_{l}(\rho)=A_{l}j_{l}(k\rho)\end{equation}  We thus reobtain
the general solution in spherical coordinates (\ref{wave}).

We can now generalize the above analysis to investigate the
behavior of GWs when they interact with a point mass $M$. In the
presence of a point mass and cosmological expansion, the
appropriate background metric is the McVittie metric. In the
Newtonian limit, using comoving coordinates the McVittie metric is
\cite{Antoniou:2016obw,Nesseris:2004wj}:
\begin{equation}\label{nmcvitttiemetric} ds^2=(1-\frac{R_s}{\rho
a(t)})dt^2-a(t)^2[d\rho^2+\rho^2(d\theta^2+\sin^2{\theta}d\phi^2)]
\end{equation}
where  $R_s=2GM$. The angular variable $\theta$ separates and thus
we use the perturbation ansatz
\begin{equation}\label{hanzmass} H_k(t,\rho,\theta,\phi)=
Q(t,\rho)P_{l}(cos\theta)
\end{equation}

\begin{figure*}[ht] \centering
\begin{center}
$\begin{array}{@{\hspace{-0.10in}}c@{\hspace{0.0in}}c}
\multicolumn{1}{l}{\mbox{}} & \multicolumn{1}{l}{\mbox{}} \\
[-0.2in] \epsfxsize=3.4in \epsffile{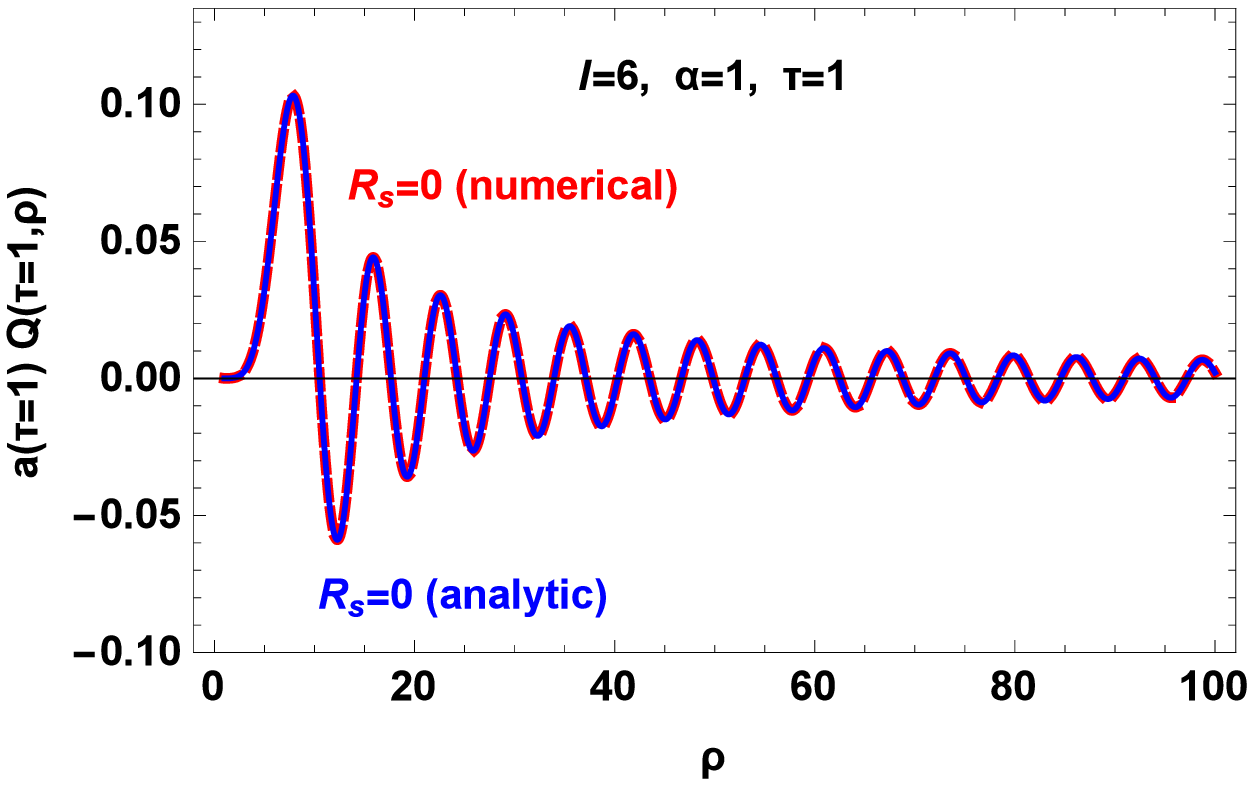} & \epsfxsize=3.4in
\epsffile{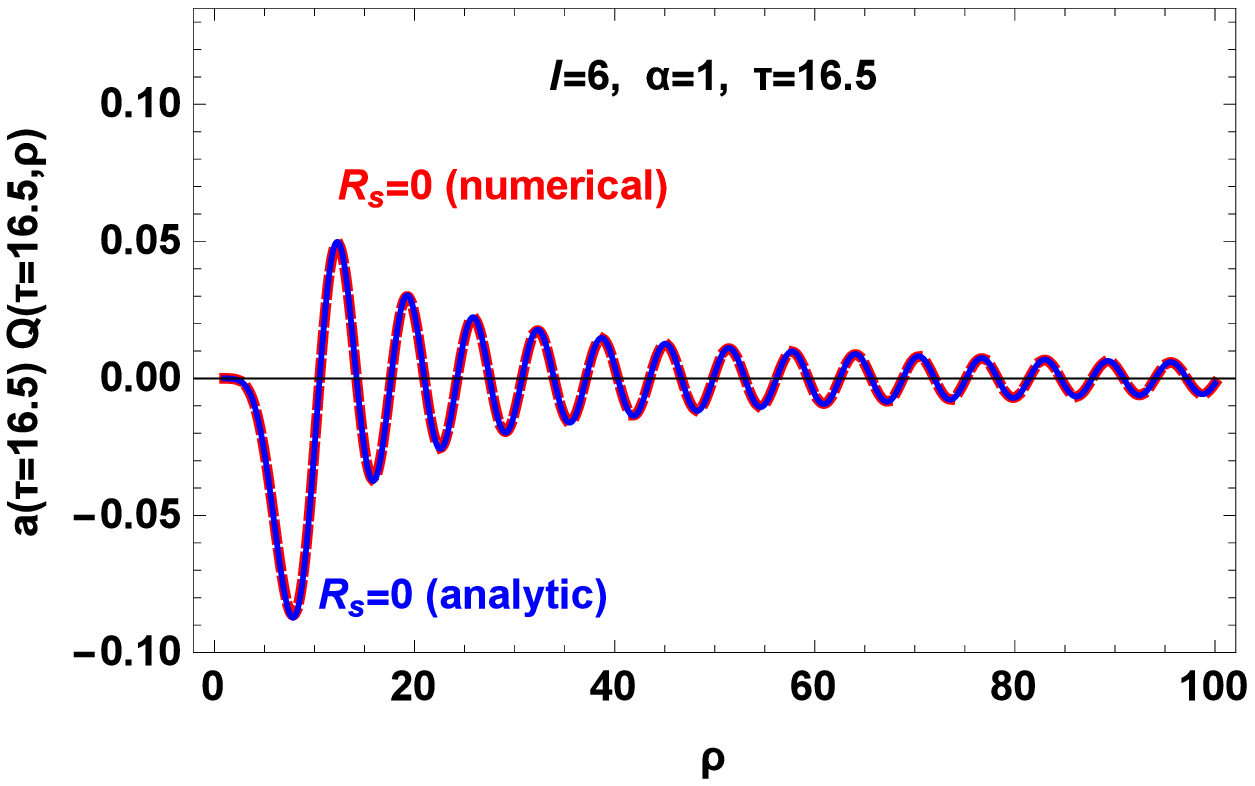} \\
\end{array}$
\end{center}
\vspace{0.0cm} \caption{\small a: A superposition of the analytic
solution with the numerical simulation initial condition taken at
$\tau=1$. The spherical wave with $l=6$ and scale factor
$a(\tau)=\tau$ is shown.   b: The numerically evolved solution (dashed red line) for
the gravitational wave $Q(\tau,\rho)$ at $\tau=16.5$ with $R_s=0$, is in
excellent agreement with the corresponding analytic evolved
solution (blue line). This is a test of the quality of the numerical
solution.} \label{fig1}
\end{figure*}

Using the background metric (\ref{nmcvitttiemetric}) and the
ansatz (\ref{hanzmass}) in the gravitational wave equation
(\ref{waveeq}) we obtain the dynamical equation for $Q(t,\rho)$ as
\begin{eqnarray}\label{qpde}
&&\frac{\partial^2 Q}{\partial \rho^2}+\frac{2}{\rho}(1-\frac{3 R_s}{4a\rho})(1-\frac{R_s}{a\rho})^{-1}\frac{\partial Q}{\partial \rho}-\frac{l(l+1) Q}{\rho^2}=\nonumber\\
&&\frac{a^2}{1-\frac{R_s}{a\rho}}[\frac{\partial^2 Q}{\partial
t^2}+\frac{3\dot{a}}{a}(1-\frac{7
R_s}{6a\rho})(1-\frac{R_s}{a\rho})^{-1}\frac{\partial Q}{\partial
t}] \end{eqnarray}
 Assuming that
\begin{equation}\label{smallrs}
\frac{R_s}{a\rho}\ll 1
\end{equation}
and keeping terms $R_s/a\rho$ only in first order we can write eq.
(\ref{qpde}) in conformal time as
\begin{eqnarray}\label{qpdeconf}
&&(1+\frac{R_s}{a\rho})\frac{\partial^2 Q}{\partial
\tau^2}+\frac{2a'}{a}(1+\frac{3R_s}{4a\rho})\frac{\partial
Q}{\partial \tau}\nonumber\\
&&=\frac{\partial^2 Q}{\partial
\rho^2}+\frac{2}{\rho}(1+\frac{R_s}{4a\rho})\frac{\partial
Q}{\partial \rho}-\frac{l(l+1) Q}{\rho^2}
\end{eqnarray}

Eq. (\ref{qpdeconf}) is not separable and it is not tractable
analytically in a simple manner. As expected, in the limit of zero
mass it separates and reduces to eqs. (\ref{rdiffeqsep}) and
(\ref{fdiffeqsep}).

In the next section we integrate eq. (\ref{qpdeconf}) numerically
and investigate the dependence of the solution on the values of
the parameter $R_s$. It will be seen that as the wave approaches
the point mass it experiences two types of distortion
\begin{itemize}
\item gravitational time delay and increase of its period in conformal cosmological time. \item
Its amplitude increases in comparison to the amplitude it would
have in the absence of the point mass. \end{itemize}

According to general relativity the expected period of the wave at
a comoving distance $\rho$ from the point mass, as measured by an
observer at infinity, is
 \be
T=\frac{T_0}{\sqrt{1-\frac{R_s}{a\rho}}}\label{periodatrho}\ee
where $T_{0}$ is the corresponding period at infinity (or in the
absence of the mass). For small mass or large distance from the
source the increase of the period is \be \frac{\Delta
T}{T_0}=\frac{1}{2a\rho}R_s\label{periodincrease}\ee where $\Delta
T=T-T_0$ is the difference of the wave periods with and without
the presence of the mass. The validity of eq.
(\ref{periodincrease}) for the GW in the vicinity of a point mass
will be demonstrated numerically in the next section.

\section{Numerical Analysis}
\label{sec:Section 3}

In order to keep the analogy with the massless case $R_s=0$ we
rescale the dynamical equation (\ref{qpdeconf}) to dimensionless
form using the wavenumber $k$ defining $k\rho=\bar{\rho}$,
$k\tau=\bar{\tau}$. In this case we have an additional physical
dimensionless parameter: \be \bar{R_s}=k R_s \label{rsdim}\ee In
the numerical analysis that follows we only use dimensionless
quantities even though we will omit the bar in what follows.

\begin{figure*}[ht]
\centering
\begin{center}
$\begin{array}{@{\hspace{-0.10in}}c@{\hspace{0.0in}}c}
\multicolumn{1}{l}{\mbox{}} & \multicolumn{1}{l}{\mbox{}} \\
[-0.2in] \epsfxsize=3.4in \epsffile{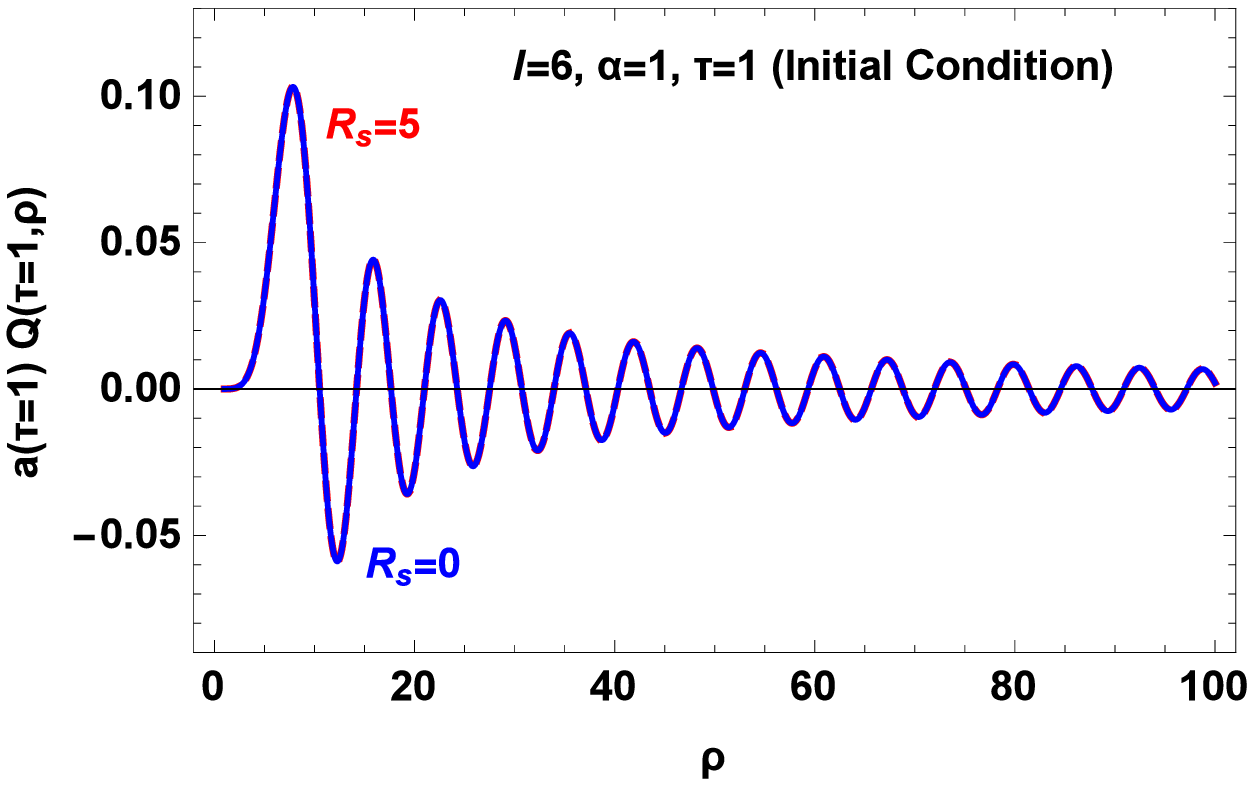} & \epsfxsize=3.4in
\epsffile{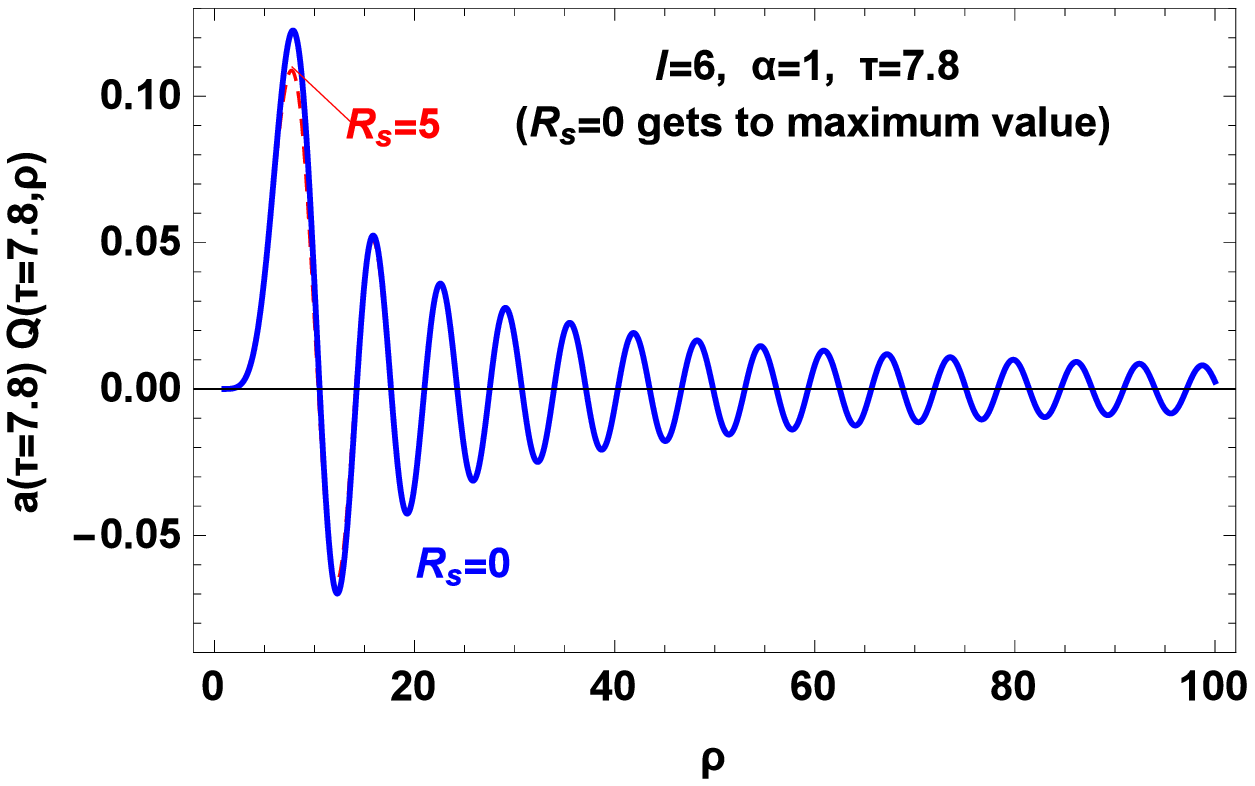} \\
\epsfxsize=3.4in \epsffile{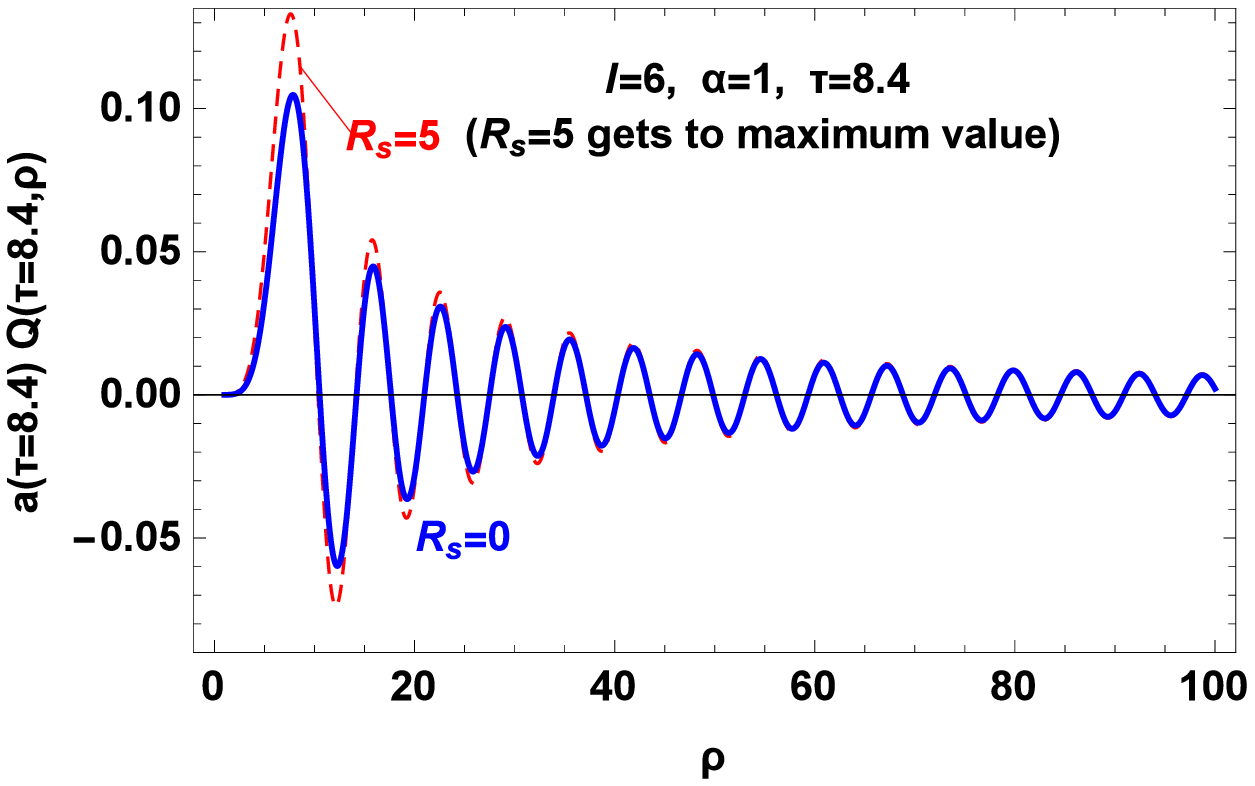} & \epsfxsize=3.4in
\epsffile{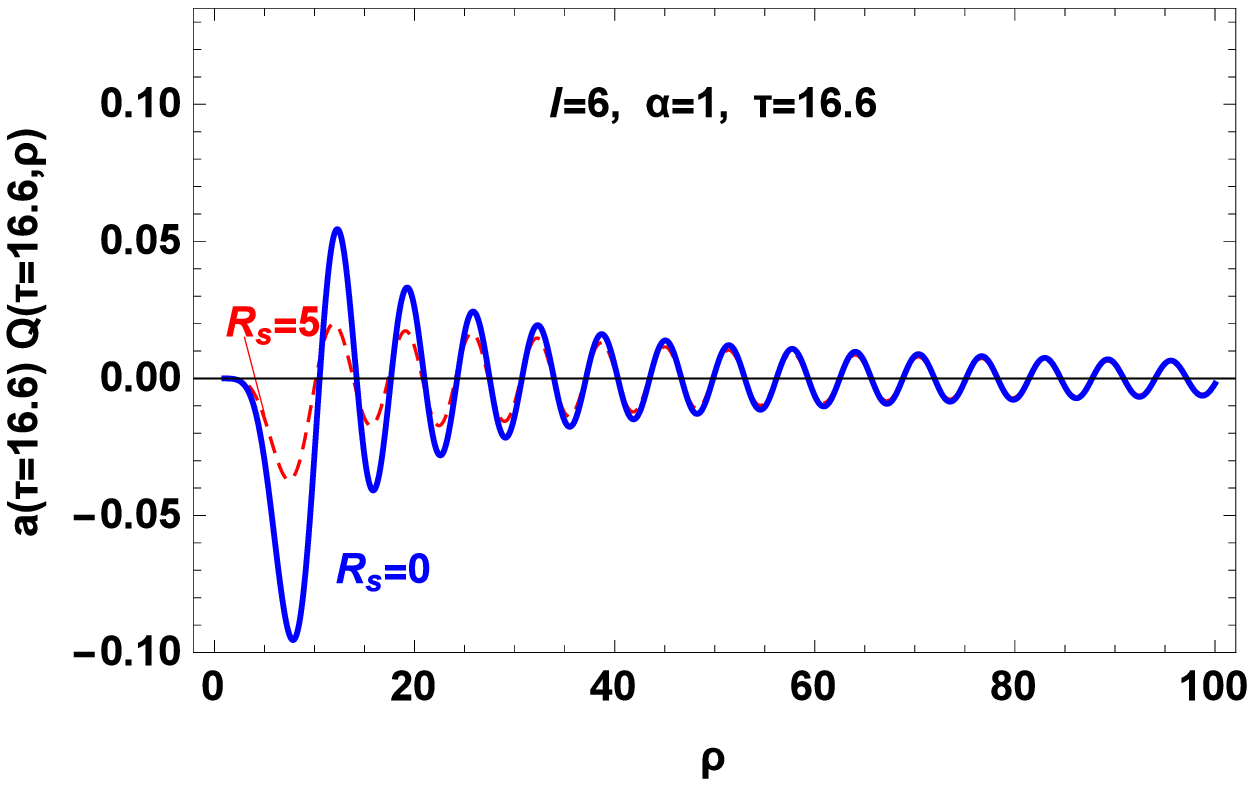}
\end{array}$
\end{center}
\vspace{0.0cm} \caption{\small The evolution of the profile of a
partial spherical gravitational wave with $l=6$, $R_s=5$ (red
dashed line) in comparison with the corresponding free solution
($R_s=0$, blue continuous line).
 The free wave reaches its maximum first (Fig. 1b) while the wave in the presence of the point mass shows a
  delay in reaching its maximum (Fig. 1c) due to gravitational redshift. The wave in the presence of the mass
  has a higher amplitude (compare 1b with 1c) in the vicinity of the mass and the phase difference increases
  with time in the vicinity of the mass (Fig. 1d).}
\label{fig2}
\end{figure*}

We solve numerically eq. (\ref{qpdeconf}) with initial conditions
corresponding to a standing gravitational wave evolving in a homogeneous
FRW spacetime ($R_s=0$) using eq. (\ref{wavesol2}) with $B_k=0$
starting the evolution at $\tau=1$. This is equivalent to assuming
that the point mass appears at $\tau=1$. For definiteness we set
$\alpha=1$ or $a(\tau)\sim \tau$ corresponding to an expanding
background in the radiation era. The boundary conditions are
imposed for $Q$ and for its first derivative at large $\rho$ where
the effects of the point mass are negligible and also correspond
to a standing GW evolving in a homogeneous FRW spacetime ($R_s=0$) using
eq. (\ref{wavesol2}) with $B_k=0$. 

The Bessel function boundary condition (\ref{wavesol2})-(\ref{wave}) we
have used for the wave equation at large distance from the source, describes a standing
GW which however can be expressed as a superposition of two propagating modes (Hankel functions). The asymptotic behaviour of Hankel functions, which is proportional to $e^{i\tau}$, corresponds to a propagating GW, while the asymptotic behavior of Bessel functions is proportional to $\cos(\tau)$ and corresponds to a standing GW.

We stress that since we have made the Newtonian approximation our results are reliable in
regions where the weak field condition (\ref{smallrs}) is
satisfied.

We thus construct numerically the solution $
Q(\rho,\tau,R_s,l,\alpha)$ and compare with the corresponding
analytical solution  $ Q(\rho,\tau,R_s=0,l,\alpha)$. We have
tested our numerical evolution by verifying that the numerical
solution for $R_s=0$ agrees with the corresponding analytical
solution at a level better that $1\%$ (Fig. \ref{fig1}).

Following the above comments about the boundary condition we fix
$Q$ (and its derivative) at the boundary $\rho_{end}>500$ using eq. (\ref{wave}) with $\alpha=1$, $A_k=1$ and $B_k=0$ as \be
Q(\tau,\rho_{bound})=\sqrt{\frac{2}{\pi}}\frac{\sin(\tau)}{\tau}j_{l}(\rho_{bound})\label{boundary}\ee
Similarly, the initial conditions set at $\tau_i=1$ are: \be
Q(\tau_i,\rho)= \sqrt{\frac{2}{\pi}}\frac{\sin(\tau_i)}{\tau_i}j_{l}(\rho)\label{initial1}\ee
and \be \frac{\partial Q(\tau_i,\rho)}{\partial
\tau}=\frac{\partial}{\partial \tau}[\sqrt{\frac{2}{\pi}}\frac{\sin(\tau)}{\tau}j_{l}(\rho)]|_{\tau=\tau_i}\label{initial2}\ee
In addition to the test of the validity of numerical solution
presented in Fig. \ref{fig1} we have performed other tests
including the verification of the independence of the numerical
solution from the location of the boundary for $\rho_{bound}>200$.

We have solved the partial differential equation (\ref{qpdeconf})
for various values of $l$ with results that are qualitatively
similar. For definiteness we present in Fig. \ref{fig2} the
solution corresponding to $l=6$ for $R_s=5$ superposed with the
corresponding solution for $R_s=0$ in order to identify the new
features introduced in the evolution of the GW by the presence of
the point mass. 

There are three main features to observe in Fig.
\ref{fig2}. First, the waves are practically identical far away
from the point mass as expected. Second, there is a time delay for
the wave in the presence and in the vicinity of the point mass
(Fig. 2b upper right). Third, the amplitude of the wave in the
presence and in the vicinity of the mass increases (compare Fig.
2b (upper right) with Fig. 2c (lower left)).

The main effect of the expansion is to reduce the amplitude of the
gravitational wave by a factor proportional to the scale factor in the absence of the
mass. This is shown in Fig. \ref{fig3} which shows that the amplitude multiplied by the
scale factor remains constant in the absence of the mass (blue oscillating line has
constant amplitude) for the particular time dependence of the scale factor considered
($a(\tau)\sim \tau$). In the presence of the mass however, the decrease of the amplitude due to
the expansion is less efficient (red line) and the product of the amplitude times the
scale factor increases slowly with time.

The gravitational wave time evolution shown in Fig. \ref{fig3} corresponds to $\rho=7.9$ (closest
maximum amplitude to the mass for $l=6$) for $R_s=5$ (red dashed
line) and is superposed with the corresponding evolution for $R_s=0$
(blue continuous line). This plot demonstrates the relative
(linear) increase of the  amplitude with time, as well as the
increased period of the wave in the presence of the mass. It also 
demonstrates (as discussed above) the well known fact that the wave amplitude in the
absence of the mass ($R_s=0$) is inversely proportional to the
scale factor (the blue wave has a constant amplitude).

\begin{figure}[!t]
\centering
\vspace{0cm}\rotatebox{0}{\vspace{0cm}\hspace{0cm}\resizebox{0.49\textwidth}{!}{\includegraphics{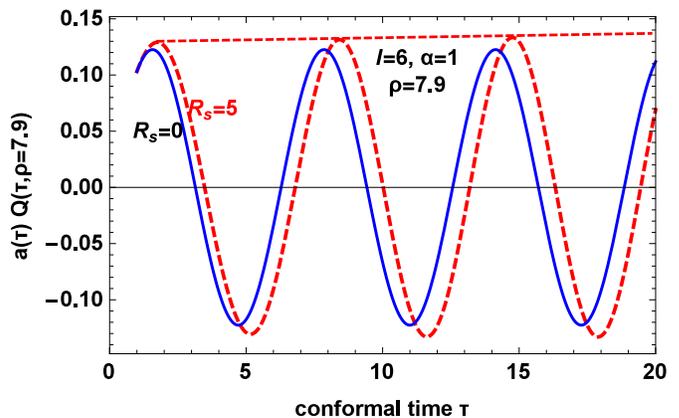}}}
\caption{ The time evolution of the first spatial maximum (at
$\rho=7.9$) of the  partial spherical wave with $l=6$, $R_s=5$
(red line)  in comparison with the corresponding free solution
($R_s=0$, blue continuous line) at the same spatial point. The
free wave reaches its maximum first (Fig. 2b) while the wave in
the presence of the point mass shows a delay in reaching its
maximum (Fig. 2c) due to gravitational redshift. The wave in the
presence of the mass has an amplitude that increases with time as
indicated with the dashed red line that in tangent to the
gravitational wave maxima. As expected, the product $a(\tau) Q(\tau)$ is constant for the free wave in an expanding background. } \label{fig3}
\end{figure}

The effects of the gravitational time delay on the evolution of
the wave may also be demonstrated by plotting the power spectrum
obtained by a Fourier series expansion of the evolving in
conformal time numerical solution at $\rho=7.9$ in harmonic waves.

The finite time interval power spectrum may be defined through the expansion
\be
Q(\tau,\rho)=\frac{a_0}{2}+\sum \limits_{i=1}^n(a_n
\cos(n\tau)+b_n\sin(n\tau))\label{coef}\ee
as
\be P(n)\equiv \log\sqrt{a_n^2+b_n^2}\label{spectrum}\ee

We used a time interval of approximately  two
complete oscillations which corresponds to a time interval
$\tau \in [1,20]$ ($\tau_i=1, \tau_{max}=20$ as shown in Fig. \ref{fig4}).

As shown in Fig. \ref{fig4} the presence of the mass (red continuous line)
leads to an increase of the amplitude of low harmonics and
decrease of the amplitude of higher harmonics which is consistent
with the effects of gravitational time delay.  The exact
form of the spectrum clearly depends on the time interval
considered, however the qualitative feature of higher amplitudes
for lower frequencies persists for all time intervals. This
feature is more prominent for lower values of $\rho$.

\begin{figure}[!b]
\centering
\vspace{0cm}\rotatebox{0}{\vspace{0cm}\hspace{0cm}\resizebox{0.49\textwidth}{!}{\includegraphics{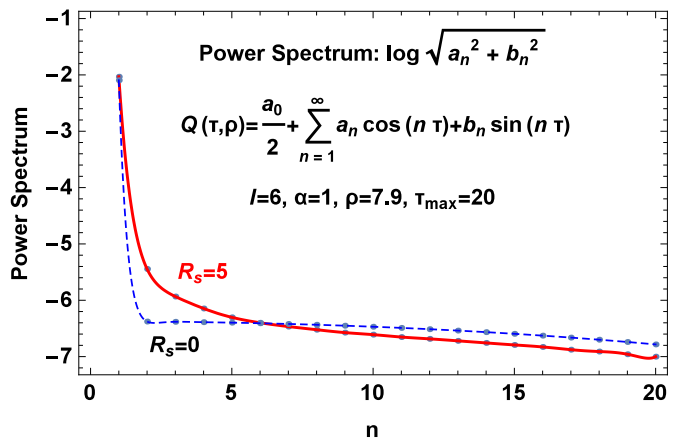}}}
\caption{The time power spectra of the gravitational wave in the
presence (red line) and in the absence (blue line) of the mass. Notice that
lower frequencies have a higher amplitude for the wave in the
presence of the mass as expected due to the gravitational time
delay.} \label{fig4}
\end{figure}

In accordance with eq. (\ref{periodincrease}) the increase of the
period of the wave at a given distance from the mass is
proportional to the mass in the weak field approximation. This is
consistent with our numerical solution as shown in Fig. \ref{fig5} where we
show the relative increase of the period of the wave $\Delta
T/T_0$ at given distances $\rho$ 
from the mass ($\rho=7.9$ and $\rho=15.89$) for various values the parameter $R_s$ (points in
plot). In order to evaluate the relative change of the period $\Delta
T/T_0$ we use the time evolution of the wave perturbation as shown in Fig. \ref{fig3} to obtain the period of the wave in the
presence of the mass and the corresponding period in the absence of the mass. Superposed in Fig. \ref{fig5} 
is the best fit straight line in each case. As
is theoretically expected there is a linear relationship in
accordance with eq. (\ref{periodincrease}). The correlation
coefficients of the points with the corresponding best fit straight line are
equal to $0.99$ indicating an excellent quality of fit.

The theoretically predicted slope is $\frac{1}{2a\rho}$ where the scale factor can
be taken as approximately constant and equal to its average value
during the wave period used to evaluate $\Delta
T/T_0$.

\begin{figure}[!t]
\centering
\vspace{0cm}\rotatebox{0}{\vspace{0cm}\hspace{0cm}\resizebox{0.49\textwidth}{!}{\includegraphics{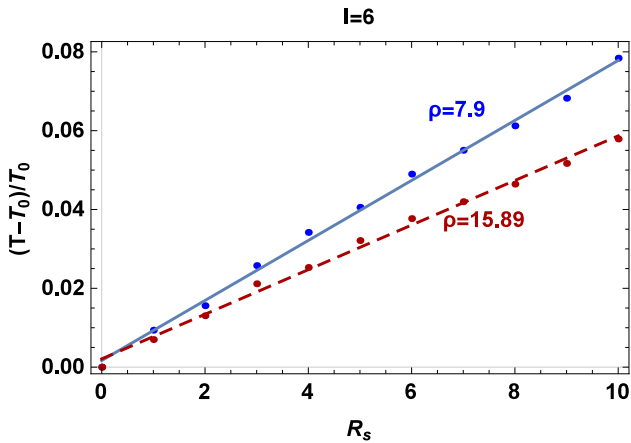}}}
\caption{The relative difference of the wave periods $\Delta
T/T_0$, where $T$ is the period in the presence of mass and $T_0$
is the period in the absence of mass, as a function of $R_s$. It
is clear that as the value of the variable $\rho$ increases, the
statistical slope of the curve decreases. This is an anticipated
result due to theoretical slope of the curve $\frac{1}{2a\rho}$.}
\label{fig5}
\end{figure}

In order to estimate the theoretical value of the scale factor, we
calculate the mean value $\bar{a}(\tau)$, in the time interval
$\tau_{1}-\tau_{2}$ of a single period, through the formula: \be
\bar{a}(\tau)=\frac{1}{\tau_{2}-\tau_{1}}\int_{\tau_1}^{\tau_2}a(\tau)d\tau=\frac{\tau_2+\tau_1}{2}\label{bara}\ee
 The observed
deviations by about $20\%$ between theoretically expected slope
and numerically obtained can be attributed to the approximations
we have made which include, the weak field assumption  ($R_s \ll
a\rho$ while in the cases considered $\frac{R_s}{a\rho} \leq
0.1$), the assumed constant scale factor for
the evaluation of the slope etc. As shown in Figs 2 and 3 the amplitude of the wave also increases
as the point mass is approached. A quantitative estimate of this
effect is shown in Fig. \ref{fig6} where we show the ratio of the
amplitudes of the waves $A/A_0$ in the presence of a mass ($A$)
and in the absence of the mass ($A_0$) for various values of the
parameter $R_s$, when $\rho=7.9$ and $\rho=15.89$. The best fit
straight line is also superposed on the points showing that a
linear relationship between $A/A_0$ and $R_s$ is a good
approximation.
\begin{figure}[!b]
\centering
\vspace{0cm}\rotatebox{0}{\vspace{0cm}\hspace{0cm}\resizebox{0.49\textwidth}{!}{\includegraphics{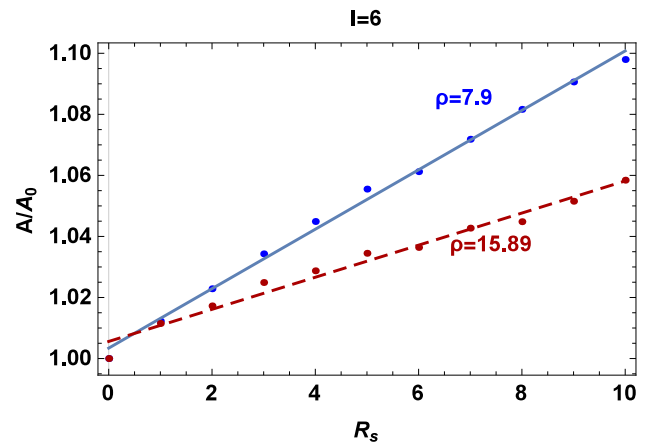}}}
\caption{The ratio of the amplitudes of the waves $A/A_0$ in the
presence of a mass ($A$) and in the absence of the mass ($A_0$) as a function of
the parameter $R_s$, when $\rho=7.9$ and $\rho=15.89$.}
\label{fig6}
\end{figure}
The amplitude increases up to $10\%$ when $\rho=7.9$ and $R_s=10$,
while for $\rho=15.89$ and the same value of $R_s$, the increase is about $5\%$. Thus the
amplitude increase appears to vary inversely proportional with
$\rho$ which is consistent with the fact that the GW gains energy
as it enters regions of space with higher curvature.

\section{Conclusion}
\label{sec:Section 4}

The effects of a point mass on a GW evolving in an expanding
universe are determined by the mass $M$ and the physical distance
$a \rho$ of the wave from the mass through the expression
$\frac{R_s}{a\rho}\equiv\frac{2GM}{a\rho}$. In the context of a
perturbative weak field analysis we have demonstrated that a point
mass tends to increase the amplitude and the period of the GW
linearly with respect to $\frac{R_s}{a\rho}$. This result is
consistent with expectations based on gravitational time delay and
energy considerations.

Even though our numerical results were presented for the special
case of a radiation dominated cosmological background ($a(\tau)\sim \tau$) and a
specific multipole component of the wave ($l=6$) we have checked that their qualitative features persist for
all multipole components and cosmological backgrounds provided that
the weak field condition (\ref{smallrs}) is respected. Thus, even though we have considered specific spherical waves in this analysis, we anticipate that our
results can also describe a plane wave when expressed as a superposition of spherical
waves.

The time slicing we considered corresponds to the coordinate
time of the particular metric we used. This coordinate time is particularly interesting
and generic as it corresponds to the proper time of a static observer located far away
from the point mass or in the absence of the point mass. This is the standard cosmic
observer whose observations are consistent with the cosmological principle. Clearly a 
different choice of time slicing would correspond to a different observer and would lead to a different metric and thus different
results.

From the results shown in Fig. \ref{fig5} and Fig. \ref{fig6}, we conclude that $T=T_0(1+\mu R_s)$ and $A=A_0(1+\nu R_s)$ where $\mu$ and $\nu$ are the slopes of the curves which are approximately equal. Thus we have demonstrated that the energy density of GWs which is proportional to $\omega^2 A^2$ has a weak dependence on $R_s$ in the context of our weak field approximation as long as the slopes $\mu$ and $\nu$ are approximately equal.

Our result has interesting implications for the calculation of the
RGW spectrum which currently assumes
\cite{Koh:2009cy,Corda:2009bx,Grishchuk:2007uz,Zhang:2008zx,Gogoberidze:2007an,Buonanno:1996xc}
a smooth homogeneous cosmological background and ignores the
presence of mass concentrations which as shown in the present
analysis would tend to modify both the magnitude and the shape of
this spectrum. A proper stochastic analysis including the effects
of mass concentrations on the relic GW spectrum is therefore an
interesting extension of the present work.

A distortion of the RGW spectrum is expected due to the presence
of point masses on various scales due to the increase of each mode amplitude and
decrease of each mode frequency. The effect will be stronger in regions of higher
mass concentrations. On scales larger than the galactic scales the role of the point
mass could be played by a galaxy while on scales of the solar system the role of the point mass could be played by a planet. 
In the solar system the effect is expected to be rather weak and beyond the sensitivity of current experiments.

An additional interesting extension could be the drop of the weak
field approximation and the use of the full McVittie metric
\cite{McVittie:1933zz} for the study of GW evolution in an
expanding background and in the vicinity of a black hole allowing
for strong gravitational field.

Even though our numerical analysis has been well tested and
provides detailed quantitative information on the GW evolution
in the presence of expansion and a point mass, an analytical
perturbative solution describing this evolution would provide
further physical insight and appears to be a tractable useful
extension of the present work.

\textbf{Numerical Analysis Files}: The mathematica files used for the production of the figures, as well as the figures may be downloaded from \href{https://drive.google.com/open?id=0B7rg6X3QljQXck5YSmQ5Rl9HeUU}{here} or upon request from the authors.

\section*{Acknowledgements}
D. Papadopoulos would like to thank the Department of Physics of
the University of Ioannina for hospitality during the period when
part of this work was in progress. We also thank K. Kleidis for useful comments.

\raggedleft
\bibliography{bibliography}

\end{document}